\documentclass[twocolumn,showpacs,preprintnumbers,amsmath,amssymb,floatfix]{revtex4}

\usepackage{graphicx}
\begin{document}

\title{Spin and charge dynamics of stripes in doped Mott insulators.}
\author{F.F. Assaad$^{1,2}$, V. Rousseau$^{3}$, F. Hebert$^{4}$,
M. Feldbacher$^{1}$ and G. G. Batrouni$^{3}$}
\affiliation{1. Institut f\"ur Theoretische Physik III, 
   Universit\"at Stuttgart, Pfaffenwaldring 58, D-70550 Stuttgart, Germany.
   \\ 2. Max Planck institute for solid state research,
   Heisenbergstr. 1, D-70569, Stuttgart, Germany
   \\3. Institut Non-Lin\'eaire de Nice, Universit\'e de Nice-Sophia
Antipolis, 1361 route des Lucioles, 06560 Valbonne, France
\\4. Theoretische Physik, Universit\"at des Saarlandes, 66041
Saarbr\"ucken, Germany}

\begin{abstract}
We study spin and charge dynamics of stripes in doped Mott insulators
by considering a two-dimensional Hubbard model with $N$ fermion
flavors. For $N =2$ we recover the normal one-band model while for $N
\rightarrow \infty$ a spin density wave mean-field solution.
For all band fillings, lattice topologies and $N= 4 n$ the model may
be solved by means of Monte Carlo methods without encountering the
sign problem. At $N=4$ and in the vicinity of the Mott insulator, the
single particle density of states shows a gap. Within this gap and on
rectangular topologies of sizes up to $30 \times 12$ we find gapless
spin collective modes centered around $\vec{q} = (\pi \pm \epsilon_x,
\pi \pm \epsilon_y)$ as well as charge modes centered around $\vec{q} = (\pm 2
\epsilon_x, \pm 2 \epsilon_y) $, $\vec{q} = ( \pm \epsilon_x, \pm
\epsilon_y)$ and $\vec{q} = (0,0)$. $\epsilon_{x,y}$ depends on the
lattice topology and doping. 
\end{abstract}

\maketitle

The understanding of the interplay between spin and charge degrees of
freedom in doped two-dimensional Mott insulators remains a challenging
issue. On the analytical front, it is straightforward to apply the
Hartree-Fock (HF) approximation but the difficulty lies in taking
quantum fluctuations into account. On the other hand numerical methods
take into account fluctuations but are limited to small clusters due
to the size of the Hilbert space, for exact diagonalization, or due to
the minus sign problem inherent to quantum Monte Carlo (QMC)
methods. In this letter, we introduce a QMC method which lies between
HF and full quantum fluctuations and apply it to the doped two
dimensional Hubbard model.  Since the $N-$flavor model we consider breaks 
$SU(N)$ spin symmetry it favors spin states. In particular in 
the vicinity of the Mott insulator (MI) we find a stripe phase   
experimentally observed in cuprates and nickelates \cite{Emery99,Buchner00}.
Our simulations reveal the dynamics of this phase. 

Our starting point is the Hamiltonian:
\begin{equation}
\label{H_N}
	H = -t \sum_{ \left( \vec{i}, \vec{j} \right) } 
	  {\bf c}^{\dagger}_{\vec{i}} {\bf c}_{\vec{j}} 
           -\frac{U}{N} \sum_{\vec{i}} \left( {\bf c}^{\dagger}_{\vec{i}} 
            {\bf \lambda} {\bf c}_{\vec{i}} \right)^2
\end{equation}
where $\vec{i}$ labels the sites of a square lattice and the first sum
runs over nearest neighbors.  The spinors $ {\bf
c}^{\dagger}_{\vec{i}} = \left( c^{\dagger}_{\vec{i},1} \cdots
c^{\dagger}_{\vec{i},N} \right) $ correspond to fermions with $N$
flavors. For even values of $N$, $\lambda_{\alpha,\gamma} =
\delta_{\alpha,\gamma} f(\alpha) $ with $f(\alpha) = 1 $ for $ \alpha
\leq N/2$ and $ -1 $ otherwise.  At $N=2$ the model reduces to the
standard $SU(2)$-spin invariant Hubbard model since the interaction is
the square of the magnetization. Away from $N=2$ the model has
an $SU(N/2)\otimes SU(N/2)$ symmetry  which becomes clear when writing
the Hamiltonian in terms of the spinors: $\left(
c^{\dagger}_{\vec{i},1} \cdots c^{\dagger}_{\vec{i},N/2} \right)$ and
$\left( c^{\dagger}_{\vec{i},N/2+1} \cdots c^{\dagger}_{\vec{i},N}
\right)$.  As $N \rightarrow \infty$ the exact solution of the model
reduces to a spin density wave mean-field approximation.

The central observation is that at $N=4n$ the minus sign problem in
QMC approach is never present regardless of the lattice topology and
band-filling.  To illustrate this and to keep the notation simple we
consider the finite temperature auxiliary field QMC approach. After
the usual decoupling of the interaction term with a
Hubbard-Stratonovich transformation, the partition function becomes $Z
= \int {\cal D} {\bf
\phi}  e^{-N S({\bf \phi})}$, where 
\begin{eqnarray}
\label{Action}
S({\bf \phi}) & = & U \int_{0}^{\beta} {\rm d} \tau 
\frac{ \sum_{\vec{i}} \Phi^2_{\vec{i}}(\tau) }{4} - \frac{1}{2} \ln {\rm Tr} \left[ T
e^{- \int_0^{\beta} {\rm d} \tau H(\tau) }\right] \\ 
H(\tau) & = & -t \sum_{ \left(
\vec{i}, \vec{j} \right) , \sigma = \pm 1}
c^{\dagger}_{\vec{i},\sigma} c_{\vec{j}, \sigma} - U \sum_{i}
\Phi_{\vec{i}} (\tau) \left( n_{\vec{i},+} -n_{\vec{i},-} \right).
\nonumber
\end{eqnarray}
Here $T$ corresponds to time ordering and  in terms of our original
fermions we can identify: $c_{\vec{i},+} = c_{\vec{i},1}$ and
$c_{\vec{i},-} = c_{\vec{i},N/2+1}$.  As $ N \rightarrow \infty$ the
integral over the fields $\phi$ is  dominated by the saddle point
configuration, $\phi^*$ satisfying $\partial S(\phi)/\partial \phi
|_{\phi = \phi^*} = 0$. A time independent solution is
$\Phi^{\star}_{i} = \langle n_{\vec{i},+} -n_{\vec{i},-} \rangle$
where the expectation value is taken with respect to the Hamiltonian
in Eq. \ref{Action}.  This is precisely the set of self-consistent
equations obtained from the mean-field decoupling: $ n_{i,+} - n_{i,-}
= \Phi^\star_i + \left[ ( n_{i,+} - n_{i,-}) - \Phi^\star_i \right]$
for the Hamiltonian of Eq. \ref{H_N} at $N=2$.

Since ${\rm Tr} \left[ T e^{- \int_0^{\beta} {\rm d} \tau  H(\tau) }\right] $ is a
real number for $ U > 0 $, it follows that $ N S(\Phi)$ is real for $N
= 4n$. This observation allows us to interpret $e^{-N S(\Phi)}$ as a
probability distribution which we sample with Monte Carlo methods.
Hence for $N=4n$ the sign problem never occurs regardless of doping
and lattice topology.  For the simulations, we have used the related
projector auxiliary field QMC algorithm which is better suited for the
study of ground state properties \cite{Assaad02}. Dynamical
information is obtained by using the Maximum Entropy method
\cite{Jarrell96}. The details of the approach will be discussed
elsewhere.

\begin{figure}[t]
\begin{center}
\includegraphics[width=.4\textwidth]{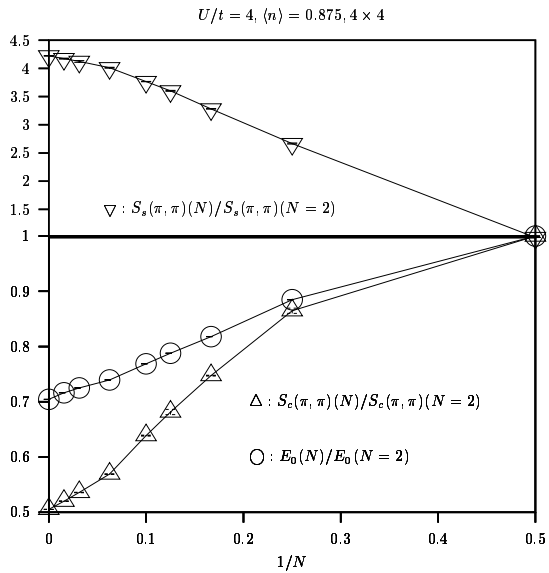}
\end{center}
\caption[]{
Ground state energy as well as spin and charge structure factors at
$\vec{q} = (\pi,\pi)$.  The data points at $N=2$ stem from exact
diagonalization studies \cite{Parola91}. }
\label{Test.fig}
\end{figure}

To test and interpret our approach, we show in Fig. \ref{Test.fig} 
the static spin and charge structure factors,
\begin{equation}
	S_{\stackrel{c}{s}}(q) =  \sum_{\vec{r}} e^{i \vec{r} \vec{q} }  
\langle (n_{0,+} \pm n_{0,-}) (n_{r,+} \pm n_{r,-}) \rangle,
\end{equation}
as well as the ground state energy as a function of $1/N$ on a $4
\times 4$ lattice, $U/t =4$ and two holes doped away from
half-filling.  The electron density is defined as: $\langle n \rangle =
\frac{1}{V} \sum_{\vec{i}} \langle n_{\vec{i},+} + n_{\vec{i},-}
\rangle $ where $V$ is the number unit cells $\vec{i}$. In spite of 
sign problems for $N \neq 4n$,  the QMC data
interpolate between the mean-field solution at $N=\infty$ and
the exact diagonalization results at $N=2$.  In the context of  the 
large-N approach \cite{Auerbach} Gaussian fluctuations correspond to 
$1/N$ corrections.   Fig. \ref{Test.fig}  shows that this approximation is 
valid till $N \simeq 16$.   Hence, $N=4$ is far beyond  this approximation. 
In particular in the strong 
coupling limit, the model at $N=4$ generates a magnetic scale proportional to
$t^2/U$; a phenomena which is beyond Gaussian fluctuations around a  weak 
coupling saddle point. 
Since we are solving the model {\it exactly} for a
given value of $N$ lattice symmetries, which are broken at the HF level,
are restored in the QMC calculations. 

\begin{figure}[t]
\begin{center}
\includegraphics[width=.5\textwidth]{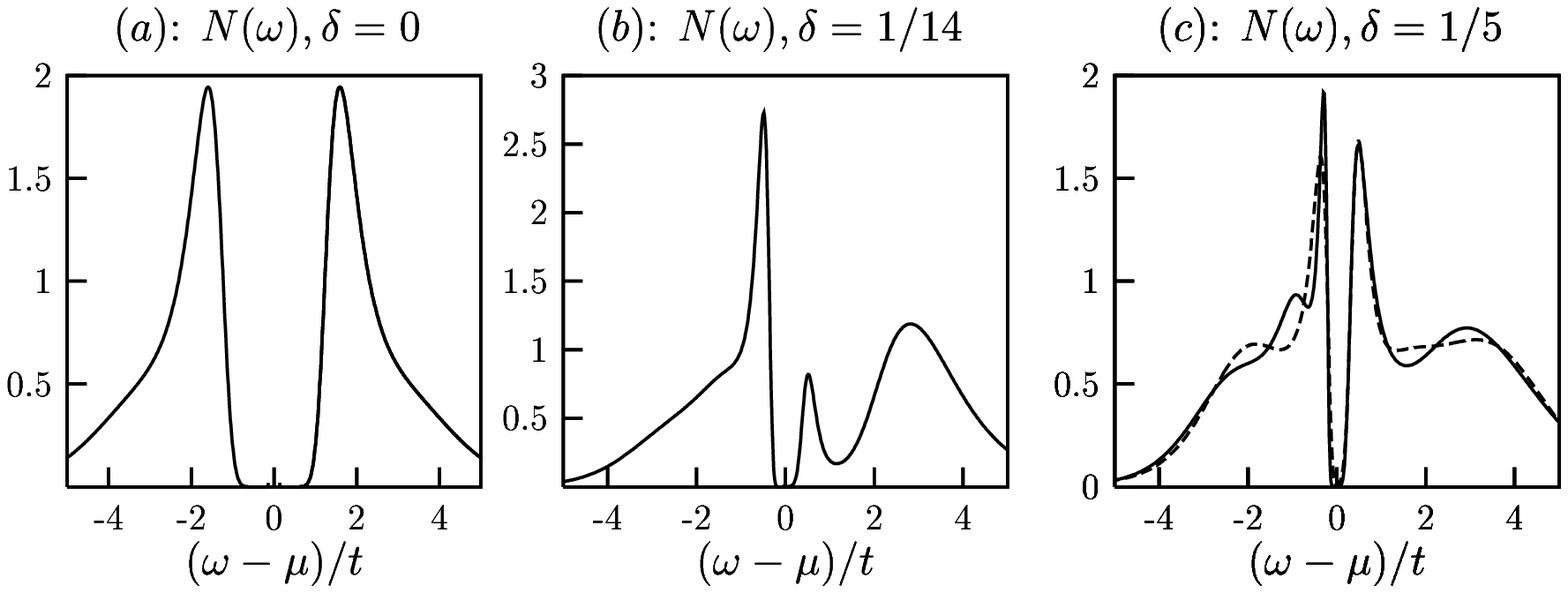} \\
\includegraphics[width=.4\textwidth, height
=0.1\textheight]{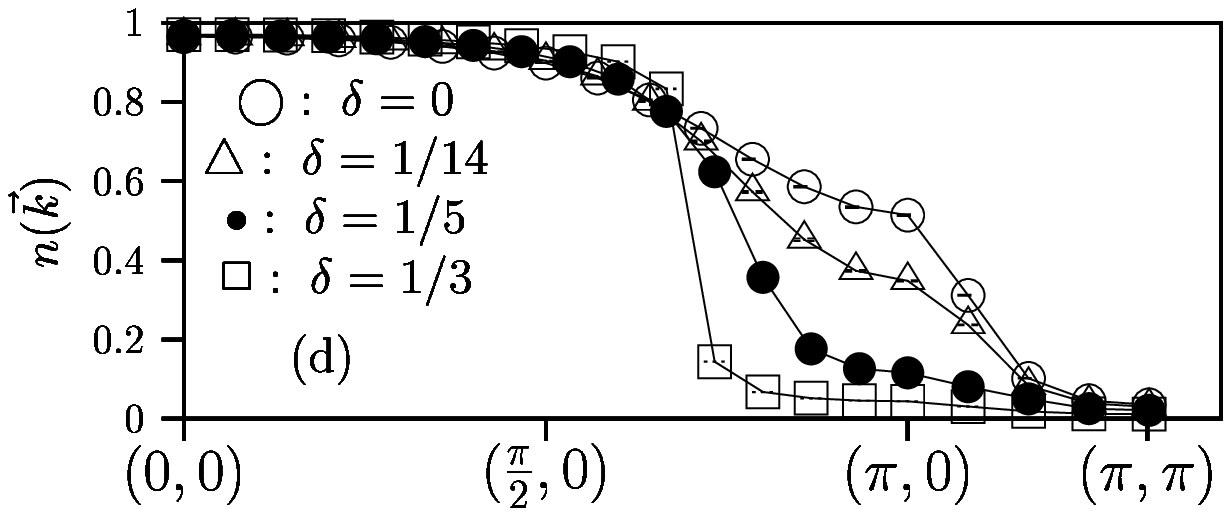} 
\end{center}
\caption[]{Density of states as a function of doping.  The lattice
sizes are: (a)-(b) $28 \times 8$, (c) $30 \times 8$ (solid line) $30
\times 12$ dashed line.  (d) Single particle occupation number. We have set $U/t =3 $,
$N=4$ and $T=0$.}
\label{Akom}
\end{figure} 

We use the method to investigate the metallic phase in the vicinity of
the MI. To maximize quantum fluctuations, we
concentrate on the $N=4$ case.  In the vicinity of the MI state, the
relevant length scale is set by the inverse doping $1/\delta$. To
achieve this length scale at least along one lattice direction, we
will consider rectangular topologies of width ranging up to 12 lattice
constants. We adopt periodic boundary conditions in both lattice directions.

We start with single particle excitations. The density of states 
$N(\omega)$ as a function of doping is plotted in Fig.
\ref{Akom} for various fillings.  At $U/t=3$,  $N =4$ and half-filling,
we see a Mott gap of approximately $2t$
(Fig. \ref{Akom}a).  Upon doping, the chemical potential shifts into
the lower Hubbard band and spectral weight is transfered from the
upper Hubbard band to generate low a energy feature
\cite{Eskes91}.  In particular, at $\delta = 1/14$ remnants of
the upper Hubbard band ($\approx 2t$ away from $\mu$) are seen and a
low energy feature is detectable. The data of Fig. \ref{Akom} also
show the presence of a small quasiparticle gap at finite dopings. This
is confirmed by the single particle occupation number $ n(\vec{k}) $
which shows a smooth behavior up to $\delta = 1/5$.  

\begin{figure}[t]
\begin{center}
\includegraphics[width=.4\textwidth, height =
0.15\textheight]{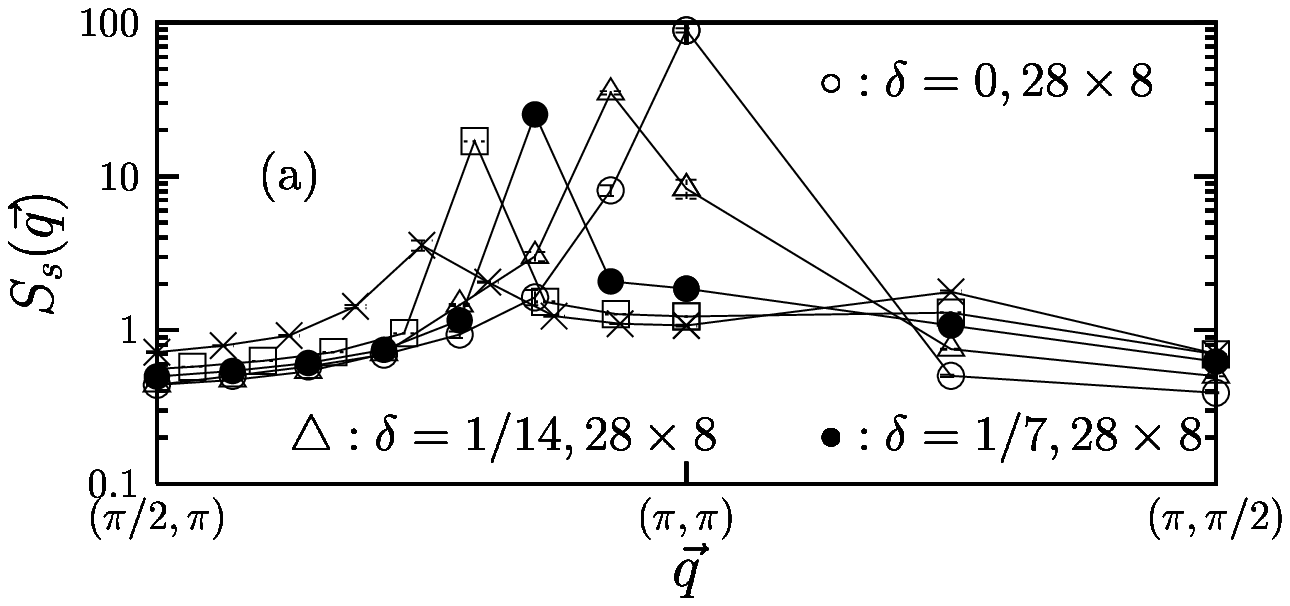} \\ 
\includegraphics[width=.4\textwidth, height =
0.15\textheight]{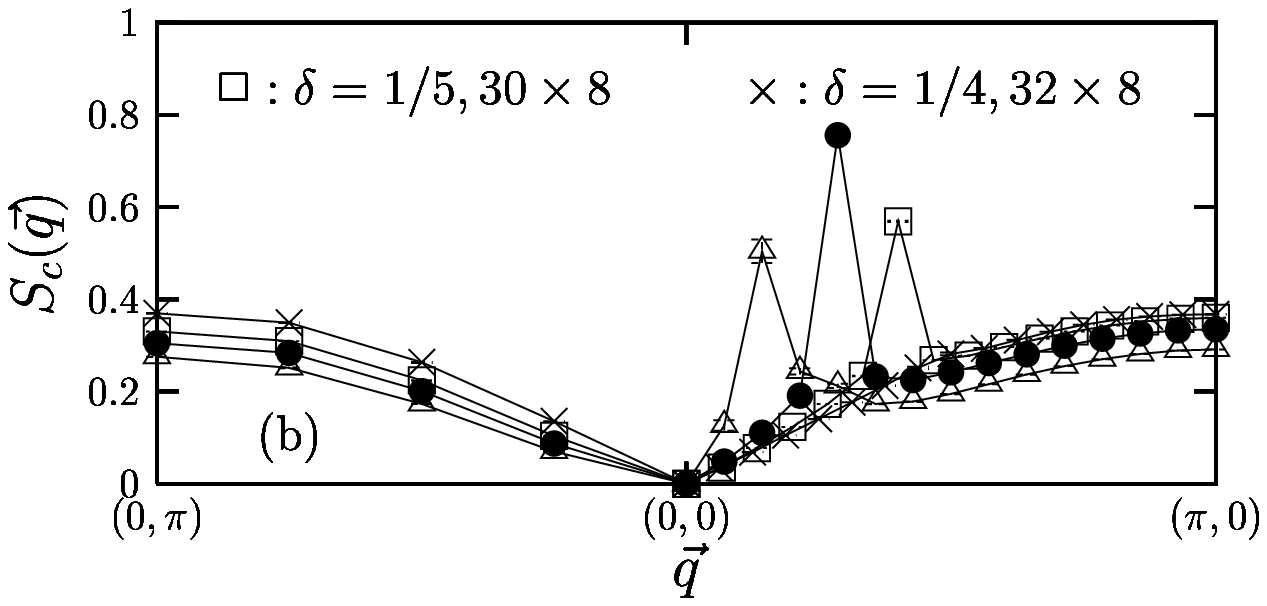} 
\end{center}
\caption[]{ Spin (a) and charge (b) structure factors as a function of
doping at $U/t = 3$.} 
\label{SpinDen}
\end{figure} 

The origin of the quasiparticle gap lies in magnetic 
fluctuations which remain substantial away from
half-filling. Fig. \ref{SpinDen} shows the spin and charge structure
factors for several dopings $\delta$ on rectangles of width $L_y = 8$.
For those topologies, the dominant features are along the $(1,0)$
direction. In particular we see a peak in the spin structure factor at
$\vec{Q}_s = (\pi - \epsilon_x, \pi) $ accompanied by one in the charge
structure factor at $\vec{Q}_c = ( 2 \epsilon_x, 0) $ with $\epsilon_x =
\pi \delta$.  Transforming the data in real space yields the stripe
caricature: rivers of holes along the $y$-direction separated by
$1/\delta$ lattice constants in the $x$-direction. Across each river
of holes, there is a $\pi$-phase shift in the spin structure with
respect to the antiferromagnetic ordering. 
Similar features are seen in  one  dimension (1D) where  our 
simulations  at $N=4$ show dominant cusps in the charge (spin) structures factors at 
$4k_f \equiv 2 \pi \delta$   ($2 k_f \equiv \pi - \pi \delta $)  \cite{Voit94}. 
In 1D we  checked successfully for power law decay of the 
$2 k_f $ spin and $ 4 k_f $ charge correlation function at  $ N=4 $. 
On fixed width  topologies  one would equally expect 
power law decay of the correlation functions. We  were however  unable to 
confirm this numerically due to limitations on lattice sizes.

\begin{figure}[t]
\begin{center}
\includegraphics[width=.4\textwidth, height =
0.13\textheight]{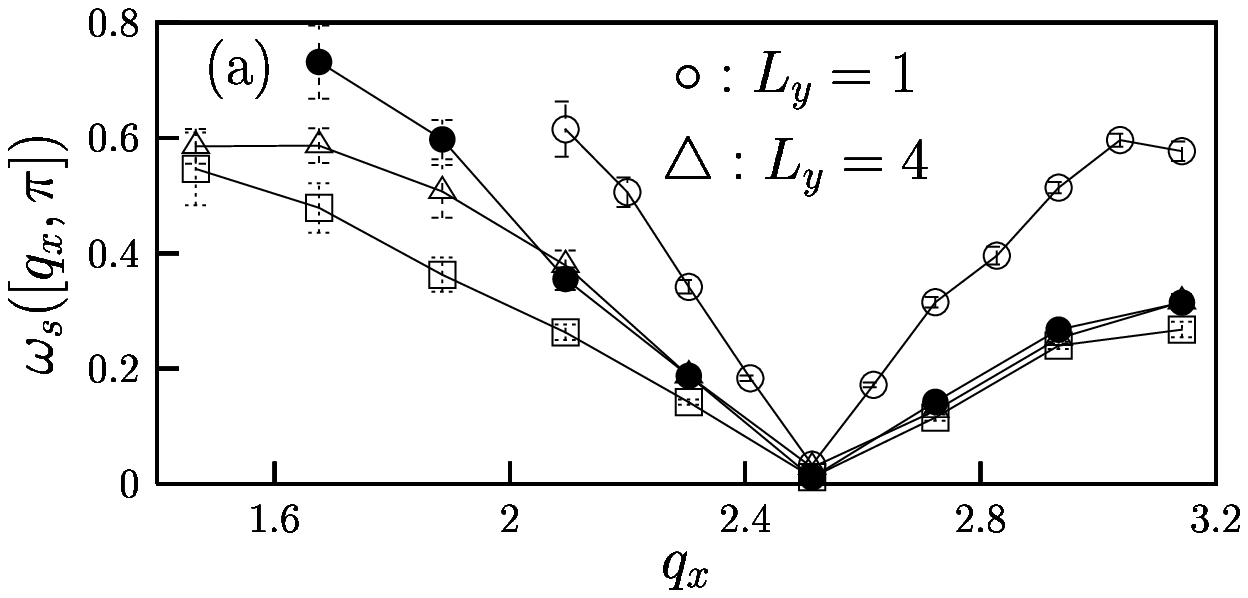} \\ 
\includegraphics[width=.4\textwidth, height =
0.13\textheight]{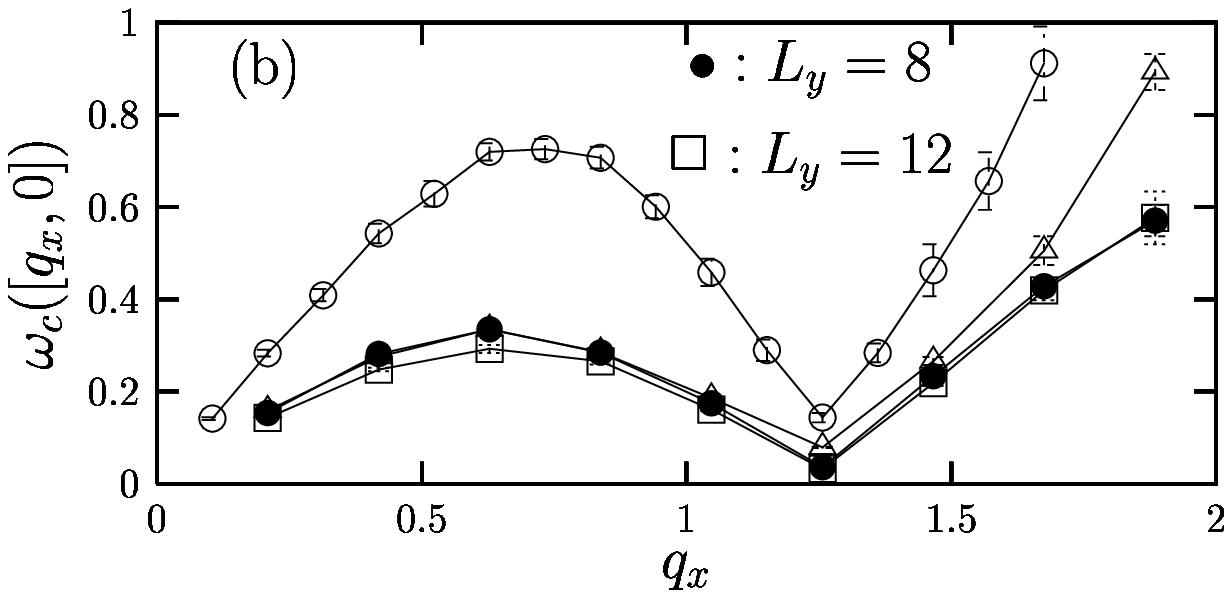} 
\end{center}
\caption[]{Leading edge of the spin (a) and charge (b) dynamical
structure factor as as function of width $L_y$ at $U/t = 3$, $\delta =
0.2$, $T=0$ and $N=4$. We consider $L_x = 60$ for $L_y=1$ and $L_x =
30$ otherwise.  Note that at $L_y = 1$ the second component of the
wave vector has no meaning. }
\label{Disp}
\end{figure} 

To study the dynamical properties of the striped phase, 
we compute the imaginary time displaced spin and charge correlation
functions : $S_{c/s} (\vec{q},\tau) $. By fitting the tail $S_{c/s}
(\vec{q},\tau) $ to a single exponential, we obtain the leading edge
of the charge and spin excitation spectra: $\omega_{c/s}(\vec{q})$.
Irrespective of lattice width the overall features of the  QMC data of 
Fig. \ref{Disp} suggest  gapless spin [charge] excitations with 
linear dispersions around $\vec{Q}_s$ 
[$\vec{Q}_c$ and $(0,0)$].  
The data equally show other features. (i) At $L_y = 8,12$, the
spin and charge excitations lie below the particle-hole continuum
which from the single particle density of states (see
Fig. \ref{Akom}(c)) lies at a threshold energy $ 2 \Delta_{qp} \approx
0.5t$. 
(ii) The continuity equation, $\frac{d }{d t}
n(\vec{x},t) + \vec{\nabla} \cdot
\vec{J}(\vec{x},t) = 0$ yields an identity between the dynamical
charge structure factor and the real part of the optical conductivity
$\sigma'(\vec{q},\omega)$ : $ \omega S_c(\vec{q},\omega) / q_x^2
\equiv \sigma'_{x,x} (\vec{q},\omega)$. 
Thus long wave length charge
modes  carry current so that we have a metallic state. 
Direct QMC evaluation of the conductivity confirms this.  

It is tempting to try to understand the above excitation spectrum 
for our finite width lattices as evolving from 1D physics   \cite{Voit94,Zaanen01}. 
Here we would however like to discuss two other possibilities in terms 
of Goldstone modes. 
(i) As mentioned previously, the model
at $N=4$ has an $SU(2) \otimes SU(2)$ symmetry and 
one may be tempted to interpret the low lying spin and charge features 
in terms of Goldstone modes associated with the breaking of this 
continuous symmetry.  Our calculations do not support this point view 
since those modes remain disordered.   
(ii)   To interpret the data in terms of the dynamics of spin and charge 
density waves \cite{Lee74,Gruener94}, we have  to argue that the pinning of those modes due to the lattice 
is small. To do so, we  carry out a HF calculation with order parameters:
$\langle n_{\vec{r},+} - n_{\vec{r},- } \rangle   = \Delta_s \cos(\vec{Q}_s \vec{r} + \phi_s) $
and
$ \langle n_{\vec{r},+} + n_{\vec{r},- } \rangle = \Delta_c \cos(\vec{Q}_c \vec{r} + \phi_c) $.
At $\delta = 0.2$ we have to $ M \vec{Q}_s = \vec{G}$ with $\vec{G} $ a reciprocal lattice 
vector and $M = 10$. The magnitude of the pinning decreases as
a function of growing values of $M$ and in the incommensurate case ($ M \rightarrow \infty $) 
vanishes.
To estimate the magnitude of the pinning, we  solve  the mean-field self-consistent equations 
for the moduli of the order parameters at fixed  values of the phases.  The mean-field energy 
then follows the form: $ E^{HF} = E^{HF}_0 + A_1 cos^2(\phi_1 + \pi/2) + A_2 cos^2(\phi_2) $  
where $\phi_1 = 2 \phi_c + \phi_s$  and $\phi_2 = - \phi_c + 2 \phi_s$. 
For the $L_y = 8$ and $L_x = 30$ system at $\delta = 0.2 $  we find that 
$A_1 \simeq  - 5 \times 10^{-5} $  while $A_2 \simeq  0.8 t $.   
This implies that the pinning of the $\phi_1$ mode
is extremely small since it is proportional to $ \sqrt{A_1} $. 
Following this point of view for the interpretation of the data, we will
have to assume that the  mass of the $\Phi_1$ mode is beyond our numerical
resolution so that we are  not  able to distinguish it from a genuine Goldstone 
mode.  Pinning $ \phi_2 = \pi/2$ and recalling that $\phi_1 = 2 \phi_c + \phi_s$, and  
$ 2 \vec{Q}_c  = \vec{Q}_s$ locks together the dynamics of the spin and charge density wave. 
The QMC results of Fig. \ref{Disp} support this point of view since the velocities of
the spin and charge modes are comparable.

\begin{figure}[t]
\begin{center}
\includegraphics[width=0.44\textwidth,height=0.2\textheight]{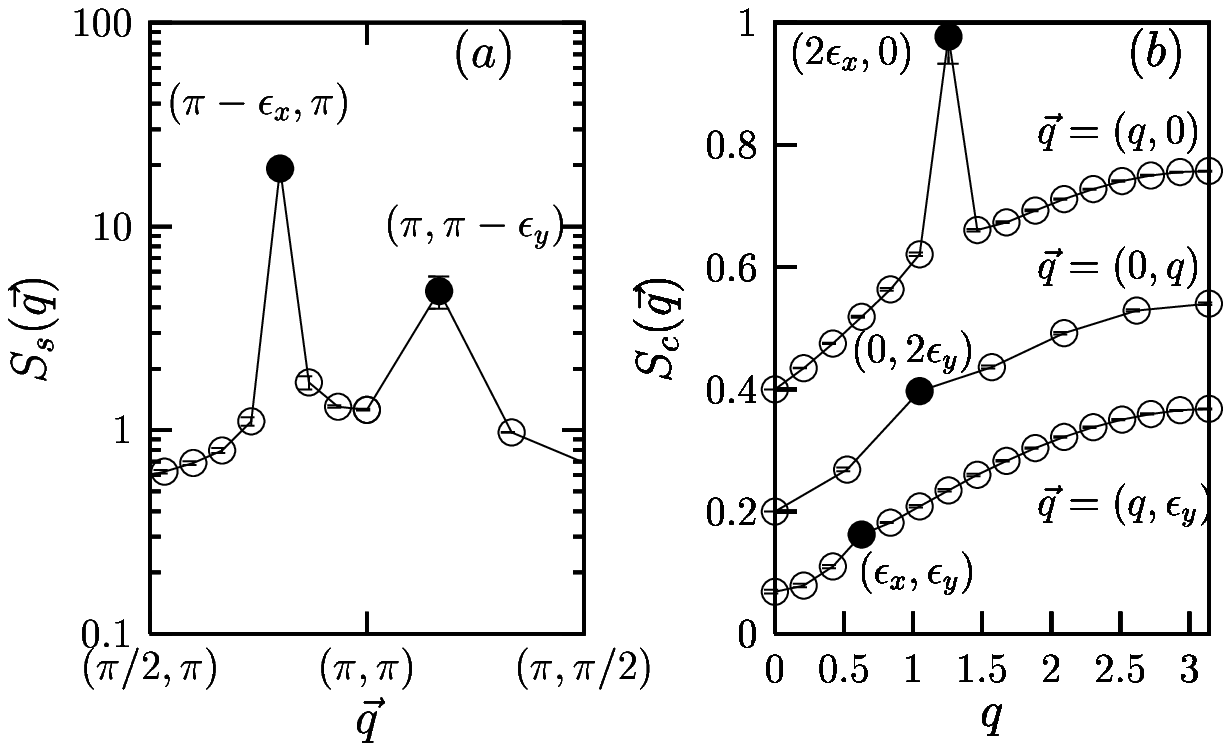}
\\
\includegraphics[width=0.22\textwidth,height=0.3\textheight]{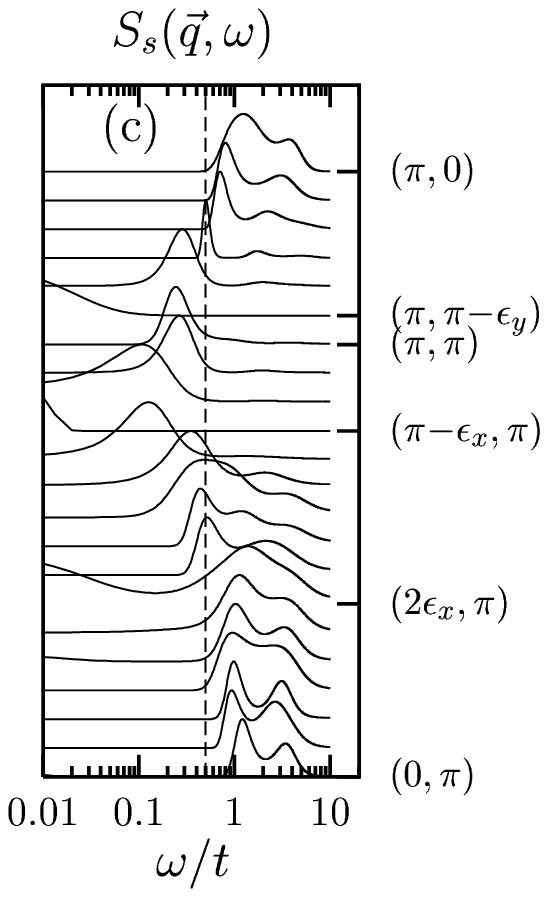}
\includegraphics[width=0.22\textwidth,height=0.3\textheight]{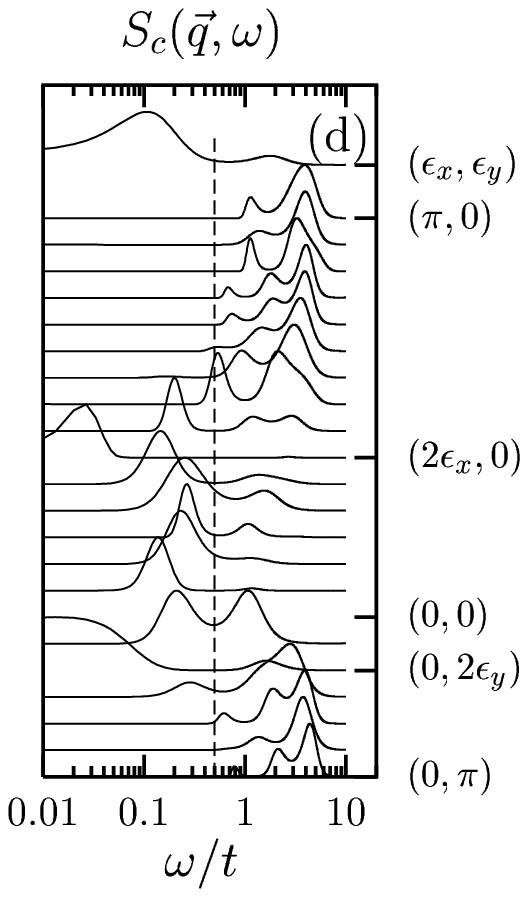}
\end{center}
\caption[]{ Ground state spin and charge degrees of freedom on a $30
\times 12 $ lattice $U/t =3 $, $\delta = 0.2$. $(a)$, $(b)$: static
spin and charge structure factors. The bullets denotes $\vec{q} $
points where cusps or peaks are seen.  $(c)$,$(d)$: dynamical spin and
charge structure factor. The dashed line corresponds to twice the
quasiparticle gap estimated from the density of states of
Fig. \ref{Akom}c.  For clarity we have normalized the data. The weight
under each curve may be recovered from Fig. (a),(b) since
$\int_{0}^{\infty} {\rm d}
\omega S_{c/s} (\vec{q},\omega) = \pi S_{c/s} (\vec{q}) $. }

\label{NSqom}
\end{figure}

We have up till now concentrated on {\it narrow} rectangular lattices
where the topology forces  stripes to occur along only  one
direction. As the width of the system increases to approach a square
topology, peaks in the structure factors are seen along both lattice
directions.  Fig. \ref{NSqom}a,b plots the spin and charge structure
factors for our largest system: $L_y = 12, L_x = 30$.  The spin shows
pronounced features not only at $ (\pi - \epsilon_x,\pi)$ but also at
$(\pi, \pi - \epsilon_y)$ where $\epsilon_x = \pi \delta$ and
$\epsilon_y = 2 \pi / L_y$. In the charge sector, cusps are seen at $
(2 \epsilon_x,0) $ and $ (0, 2 \epsilon_y)$. There is however another
feature namely a cusp at the wave vector $(\epsilon_x, \epsilon_y)$
(Fig. \ref{NSqom}b).  For this lattice size the dynamical spin,
$S_s(\vec{q},\omega)$, and charge, $S_c(\vec{q},\omega)$, structure
factors are plotted in Fig. \ref{NSqom}c,d. At the locations of peaks
in the static structure factors soft modes in the dynamics are present.  The
lower edge of the spectra along the x-direction is plotted in
Fig. \ref{Disp}.  On this lattice topology, it is possible to arrange
the holes in two stripes along the x-direction (corresponding to an
incommensurate filling) or in five stripes along the y-direction. If
the stripe ordering turns out to be long-ranged, one expects one of
the two directions to be spontaneously chosen in the thermodynamic
limit.  On the other hand if there is no long-range stripe order, just
short range fluctuations, we expect this state to remain stable as the
size of the system grows.  We note that a slave boson approach to the
$t$-$J$ model yields an identical pattern of spin and charge modes.
For given soft spin modes at wave vectors $\vec{q} = (\pi \pm
\epsilon, \pi \pm \epsilon)$ charge modes  are
generated at wave vectors $\vec{q} = ( \pm 2 \epsilon, \pm 2
\epsilon)$ and $\vec{q} = (\pm \epsilon, \pm \epsilon)$ \cite{Horsch_un}.
Our findings support this point of view.

To summarize, we have introduced a new QMC approach which allows us to
obtain insight into doped Mott insulators.  It is based on the
observation that when the number of fermion flavors $N=4n$, a broken
spin-symmetry Hubbard model may be simulated with no sign problem
regardless of the lattice topology and band filling.  More generally, 
as a function of growing values of $N$ the sign problem becomes less and 
less severe. 
Our approach clearly provides a way to go beyond Gaussian fluctuations around
a given saddle point.  On the other hand, it is at present not clear if the 
extrapolation from finite $N$ to $N=2$ is justified.
We have used this method to investigate the doped Mott
insulator at $N=4$.  We find a metallic state with no quasiparticles since
there is a gap or pseudogap  in the single particle density of states.
The low energy excitations are collective spin and charge modes.
On our largest rectangular topologies they  are {\it gapless } in
the spin sector at $( \pi \pm \epsilon_x,\pi)$ and $(\pi, \pi \pm \epsilon_y)$ 
and at $( \pm 2 \epsilon_x,0)$, $(0, \pm 2 \epsilon_y)$,
$(0,0)$ as well $(\pm \epsilon_x, \pm \epsilon_y)$ in the charge
sector.  

We wish to thank the HLR-Stuttgart for generous allocation of computer
time, the DFG for financial support (grant numbers AS 120/1-3 , AS
120/1-1 ) as well as a joint Franco-German cooperative grant
(PROCOPE). FFA thanks P. Horsch and R. Zeyher for discussions.

%\bibliographystyle{./prsty}
%\bibliography{/home/assaad/TeX/fassaad}

\begin{thebibliography}{10}

\bibitem{Emery99}
V. Emery, S. Kivelson, and J. Tranquada, Proc. Natl. Acad. Sci. USA {\bf 96},
  8814  (1999).

\bibitem{Buchner00}
H. Klauss, W. Wagener, M. Hillberg, W. Kopmann, H. Walf, F. Litterst, M.
  H\"ucker, and B. B\"uchner, Phys. Rev. Lett. {\bf 85},  4590  (2000).

\bibitem{Assaad02}
F.~F. Assaad,  in {\em Lecture notes of the Winter School on Quantum
  Simulations of Complex Many-Body Systems :From Theory to Algorithms.}, edited
  by J. Grotendorst, D. Marx, and A. Muramatsu. (Publication Series of the John
  von Neumann Institute for Computing, 2002), Vol.~NIC series Vol.
  10., pp.\ 99--155.

\bibitem{Jarrell96}
M. Jarrell and J. Gubernatis, Physics Reports {\bf 269},  133  (1996).

\bibitem{Auerbach}
A. Auerbach, {\em Interacting electrons and quantum magnetism.}, {\em Graduate
  texts in contemporary physics} (Springer, New York, Berlin, Heidelberg,
  1994).

\bibitem{Parola91}
A. Parola, S. Sorella, M. Parrinello, and E. Tosatti, Phys. Rev. B {\bf 43},
  6190  (1991).

\bibitem{Eskes91}
H. Eskes, M. Mcinders, and G.~A. Sawatzky, Phys. Rev. Lett. {\bf 67},  1035
  (1991).

\bibitem{Voit94}
J. Voit, Rep. Prog. Phy. {\bf 57},  977  (94).

\bibitem{Zaanen01}
J. Zaanen, O. Osman, H. Kruis, Z. Nussinov, and J. Tworzydlo, Philos. Mag. B
  {\bf 81},  1485  (2001).

\bibitem{Lee74}
P. Lee, T. Rice, and P. Anderson, Solid State Commun. {\bf 14},  703  (1974).

\bibitem{Gruener94}
G. Gr\"uner, Rev. Mod. Phys. {\bf 66},  1  (1994).

\bibitem{Horsch_un}
P. Horsch and G. Khaliullin, unpublished  .

\end{thebibliography}

\end{document}